# Intelligent optical performance monitor using multi-task learning based artificial neural network

**ZHIQUAN WAN, ZHENMING YU,* LIANG SHU, YILUN ZHAO, HAOJIE ZHANG AND KUN XU**

*State Key Laboratory of Information Photonics and Optical Communications, Beijing University of Posts and Telecommunications, Beijing, 100876, China*
*\*yuzhenming@bupt.edu.cn*

**Abstract:** An intelligent optical performance monitor using multi-task learning based artificial neural network (MTL-ANN) is designed for simultaneous OSNR monitoring and modulation format identification (MFI). Signals' amplitude histograms (AHs) after constant module algorithm are selected as the input features for MTL-ANN. The experimental results of 20-Gbaud NRZ-OOK, PAM4 and PAM8 signals demonstrate that MTL-ANN could achieve OSNR monitoring and MFI simultaneously with higher accuracy and stability compared with single-task learning based ANNs (STL-ANNs). The results show an MFI accuracy of 100% and OSNR monitoring root-mean-square error of 0.63 dB for the three modulation formats under consideration. Furthermore, the number of neuron needed for the single MTL-ANN is almost the half of STL-ANN, which enables reduced-complexity optical performance monitoring devices for real-time performance monitoring.

## 1. Introduction

With the explosive growth of traffic induced by internet of things (IOTs), artificial intelligence-intensive services and fifth-generation (5G) services, the capacity of optical communication system is increasing dramatically. To meet the large number of traffic demands, advanced optical modulation formats with high spectral efficiencies were widely studied [1]. Furthermore, reconfigurable optical add-drop multiplexers (ROADMs) together with flexible transceivers and software defined network (SDN) controllers enable elastic and reconfigurable optical network to meet a better utilization of available transmission capacity [2]. In such a network, it is essential to monitor various network performance parameters to optimize resources utilization and allocate adequate system margin [3]. Among various parameters of optical performance monitoring (OPM), optical signal-to-noise ratio (OSNR) is one of the most importance since it is directly related to bit-error ratio (BER). It also plays a vital role for fault diagnosis as well as links health detection. Existing OSNR monitoring techniques include error vector magnitude (EVM) [4], asynchronous delay-tap plots (ADTPs) [5,6], stokes parameters [7] and intermediate frequency analysis [8].

Besides OSNR monitoring, modulation format identification (MFI) has drawn a great interest with the development of flexible transceivers and elastic optical networks (EONs) [9]. The variation of transmission parameters in EONs imposes the adaption of digital signal processing (DSP) algorithm [10]. For example, some essential equalization algorithms used in the receiver must be suitable for the modulation format of the received signals [11]. By implementing MFI, reconfigurable DSP flow can be deployed when different advanced modulation formats signals are transmitted. In this way, modulation formats can be adjusted adaptively based on channel condition and traffic demands, which makes the optical network software-programmable [12]. Amplitude histograms (AHs) [13], ADTPs [5] and digital frequency-offset [14] can be selected as metrics for MFI.

Recently, with the leap-forward development of computing resource, deep learning (DL) has achieved great success in the area of computer vision (CV), natural language processing (NLP), recommend system [15,16] etc. As a current research hotpot, several DL architectures are applied in OPM to improve the monitoring accuracy. Convolutional neural networks (CNNs), which achieve tremendous successes in CV area, were adopted for OSNR monitoring and MFI in [17,18]. Eye diagrams and constellations were selected as processing objectives for intensity modulation with direct detection (IM/DD) system and coherent detection system, respectively. In [19], CNN combined with asynchronously sampled data successfully achieved OSNR estimation. In [12], MFI was achieved by using CNN combined with parameters in 2D Stokes planes. In [20], long short-term memory (LSTM) network was deployed for OSNR monitoring. In [13], AHs of direct detected signals was selected as input features of artificial neural network (ANN) for MFI. In [10], the authors extended the work in [13]. AHs after constant module algorithm (CMA) were used to monitor OSNR and identify modulation format simultaneously with four ANNs. One was employed for MFI purpose and other three were employed to estimate OSNR for individual modulation formats (QPSK, 16QAM, 64QAM).

In this paper, we propose a novel intelligent optical performance monitor using multi-task learning (MTL) based ANN. MTL is an approach to improve model generalization by using the information contained in the related tasks as an inductive bias [21]. It is widely used in CV area to improve pattern identification accuracy [22]. Apart from ANN, this MTL based method can also be applied in other DL architectures like CNN and LSTM. By employing this MTL-ANN in conjunction with signals' AHs after CMA, simultaneous OSNR monitoring and MFI are achieved. Experimental results obtained from 20 GBaud non-return-to-zero on-off keying (NRZ-OOK), 4-ary pulse amplitude modulation (PAM4) and 8-ary pulse amplitude modulation (PAM8) signals show that MTL-ANN can simultaneously realize OSNR monitoring and MFI with higher accuracy and stability, compared with single-task learning based ANNs (STL-ANNs). The results show an MFI accuracy of 100% for the three modulation formats under consideration. Besides, OSNR monitoring with root-mean-square error (RMSE) of 0.63 dB is achieved. Furthermore, the neurons needed for MTL-ANN is almost the half of STL-ANN. In this intelligent optical performance monitor, only a single MTL-ANN with reduced network structure is deployed which enables reduced-complexity OPM devices for real-time performance monitoring.

## 2. Operating principle

In this paper, CMA equalization is adopted to compensate linear impairments in our system since it works on modulation format unassisted mode [23]. AHs with 200 bins for three signal types considered in this work are shown in Fig. 1 for three different OSNRs. It is clear from Fig. 1 that the AHs depend on modulation format as well as OSNR, thus AHs can be exploited for simultaneous OSNR monitoring and MFI. Occurrences at each bin of AHs are selected as the input features for ANN as shown in Fig. 2. The neuron numbers in input layer are equal to the bin numbers. In Fig. 2, each circle represents a neuron and it can be modeled as a logistic unit. The output of $r$-th neuron can be expressed as:

$$A_r(x) = f\left(\sum_i w_{ri} x_i\right) \quad (1)$$

Where $x_i$ is the $i$-th input to the neuron, $w_{ri}$ is the corresponding weight for the $r$-th neuron and $f(x)$ is the activation function of the neuron. ANN model the nonlinear problem with the help of multi-neurons architecture and nonlinear activation function in each neuron. In this paper, tangent sigmoid (Tanh-sigmoid) function is selected as activation function for neurons in hidden layers. For output layer 1, which focus on MFI, softmax function is selected as activation function. The output of softmax function is a categorical probability distribution, which reveals the probability that any of the classes are true. For example, a three-output ANN with a column vector $h_w(x)=[0.8, 0.15, 0.05]^T$ means that the output has 80 percent probability belong to the first class. For output layer 2, which focus on OSNR monitoring, linear function is selected as activation function. Tanh-sigmoid function and softmax function is expressed as Eq. (2) and Eq. (3), respectively.

$$\text{Tanh}(x) = \frac{e^x - e^{-x}}{e^x + e^{-x}} \qquad (2)$$

$$\text{Softmax}(x_i) = \frac{e^{x_i}}{\sum_i e^{x_i}} \qquad (3)$$

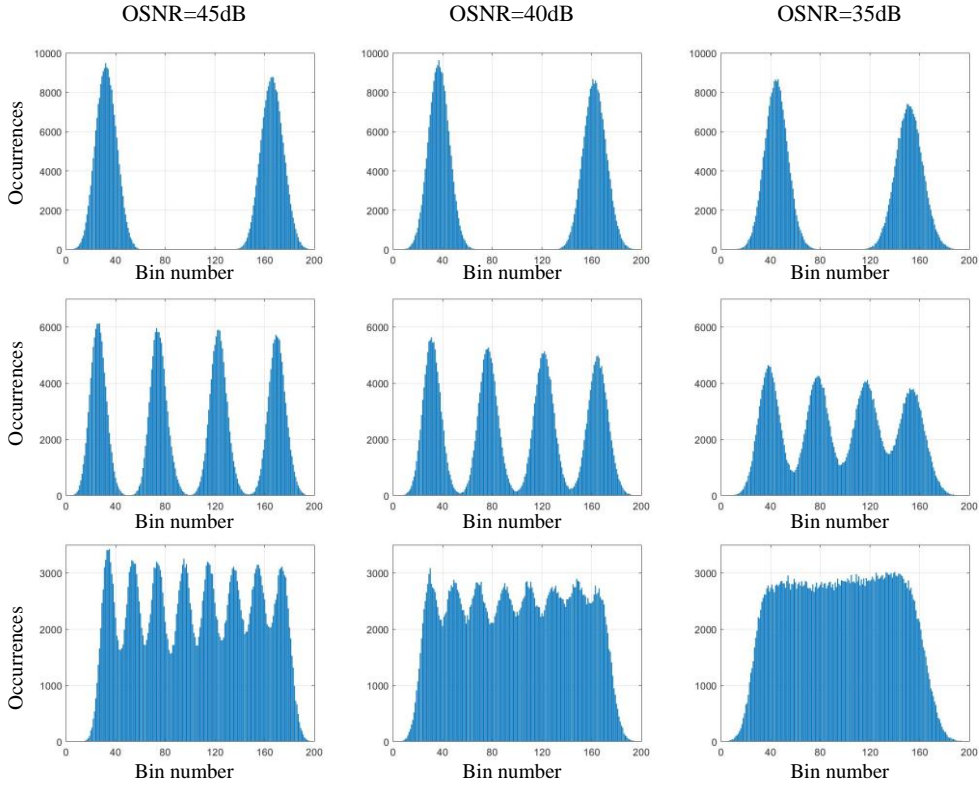

Fig. 1. AHs with 200 bins for three different OSNRs for NRZ-OOK (first row), PAM4 (second row) and PAM8 (third row) signals after CMA equalization.

For an inductive learner (neural network is an inductive learner) trained by finite samples, inductive bias can cause it to prefer some hypotheses than other hypotheses. MTL uses the training samples of related tasks as an inductive bias, so the inductive learner is biased to prefer hypotheses that have utility across multi-tasks [21]. Different with STL-ANN, MTL-ANN has multi-output layers for multi-tasks as shown in Fig. 2. Since multi-tasks share a common hidden layer, the compute resources are reduced and the commonality of different tasks can be discovered at same time. With the help of specific hidden layers for different tasks, the characteristics of different tasks are learned. The number of hidden layers and neurons in each layer can be designed for specific task. In this paper, we define the total loss function of MTL-ANN as the weighted sum of different tasks' loss function as Eq. (4) shows. In this way, effect of different tasks to the network structure is considered.

$$J(w) = \sum_{t=1}^{T}\left\{w_t \sum_{k=1}^{K_t}\left|[y_t(n)]_k - [h_t(x(n))]_k\right|^2\right\} \tag{4}$$

Where $T$ represents the number of tasks, $K_t$ represents the number of classes in $t$-th task, $[h_t(x(n))]_k$ is the output of MTL-ANN belongs to class $k$ of $t$-th task and $[y_t(n)]_k$ is the reference value of $x(n)$ which belongs to the class $k$ of $t$-th task. To consider the effect of different tasks to the network, $w_t$ is set as the loss weight of $t$-th task. In this paper, MFI is selected as task 1, OSNR monitoring is selected as task 2, so $T=2$, $K_1=3$, $K_2=1$. After network structure selected, Adam algorithm is adopted to adjust weights in MTL-ANN. Compared to traditional stochastic gradient descent algorithm, Adam algorithm is computationally efficient and requires little memory [24].

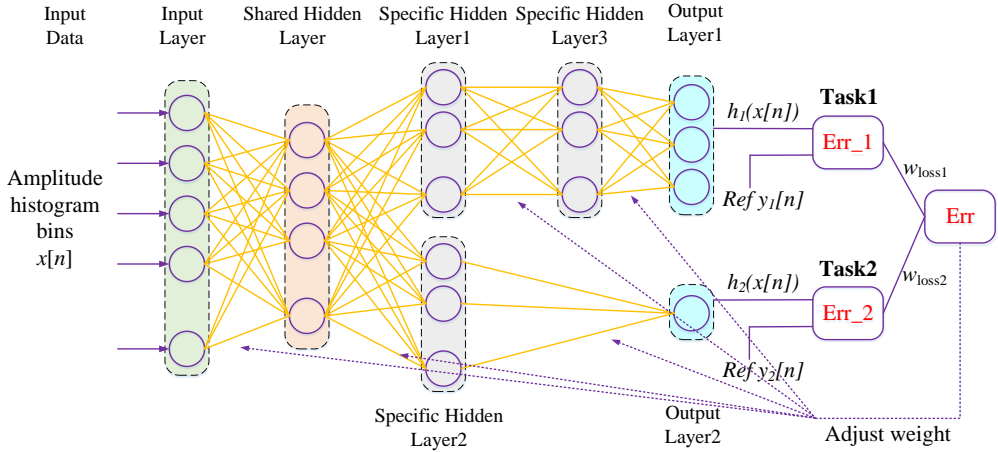

Fig. 2. Structure of MTL-ANN.

## 3. Experimental setup

The experimental setup is shown in Fig. 3. Three widely-used optical signals in IM/DD system (NRZ-OOK, PAM4, PAM8) with a pattern length of $2^{13}-1$ symbols at 20 GBaud are generated by a 50 GSa/s arbitrary waveform generator (AWG, Tektronix AWG70001A). The signal with a peak-to-peak voltage of ~1V drives the Mach-Zehnder modulator (MZM). The generated optical signal is launched to a spool of dispersion-uncompensated standard single mode fiber (SSMF) with 2.5-km fiber reach. The launched power is set as 5 dBm. After fiber transmission, an Erbium-doped fiber amplifier (EDFA) and a variable optical attenuator (VOA) are employed to load optical noise and adjust the OSNR from 32 dB to 45 dB at step of 1 dB. At the receiver, the signal is passed through a 1-nm optical band pass filter (OBPF). The resulting OSNR is measured by an optical spectrum analyzer (OSA). After detected by a PIN photodetector, the electrical signal is sampled by a 100 GSa/s digital sampling oscilloscope (DSO, Tektronix MSO72004C). The reference clock output port of AWG and the external reference clock input port of DSO are connected to synchronize the clock.

At the beginning of the offline digital signal processing (DSP) flow, we first remove the direct-current (DC) offset. After that, the data stream is resampled to two samples per symbol to enable the proposed equalization algorithm. After CMA-based equalization, linear transmission impairments is compensated. The AHs with specific bins are generated from the equalized samples. The occurrences at each bin are selected as input features for the following ANN. To investigate the efficiency of MTL-ANN, STL-ANN with identical network structure is employed as a comparison. For example, when dealing with task 1, specific hidden layer 2 and output layer 2 in Fig. 2 are discarded. After obtaining OSNR information, fault diagnosis and links health detection can be deployed. On the other hand, MFI information can be used for modulation format assisted equalization algorithm, like decision-directed least-mean-square (DD-LMS) algorithm based volterra nonlinear equalizer (VNLE) [11], to improve system BER performance. In this paper, Keras library combined with Tensorflow backend are selected as the model of ANN [25].

Based on the above system, we collect 10 AHs for each OSNR value of each modulation format. The entire data set comprises 420 (10 × 14 × 3) AHs in total. The AHs in this data set are divided into training and testing sets by random selecting 95% (i.e. 399) and 5% (i.e. 21) of all AHs, respectively. In the training sets, 10% of the data (i.e. 39) are selected as cross-validation set.

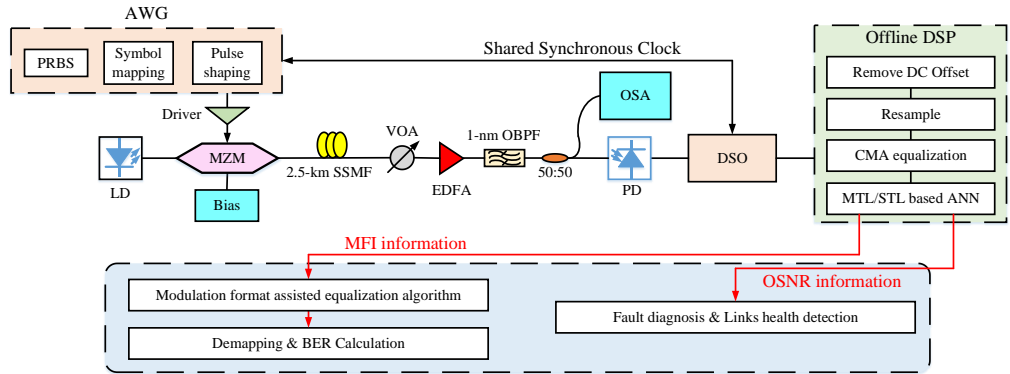

Fig. 3. Experimental setup for simultaneous OSNR monitoring and MFI.

## 4. Results and discussion

At first, we optimize the bin number of AHs. AHs with different bins for the three signal types considered in this work at 45 dB OSNR are shown in Fig. 4. It is seen that AHs with more bins can describe the received signal after CMA equalization more precisely. However, network structure will expand with the increase of bin number which may cause overfitting. In our work, cross-validation set is deployed to solve this problem. When investigating the effect of bin number, the neurons in shared hidden layer are designed a half of neurons in input layer. The neurons in specific hidden layer 1, 2 and 3 are designed a half of the neurons in their front layer, respectively. This is a common way to select neuron number in hidden layers. The ratio of OSNR loss weight to MFI loss weight is set as 1 while investigating the hyperparameters (i.e. number of hidden layers, neuron numbers in each layer) of ANN.

The MFI accuracy and OSNR estimated error versus bin number for MTL-ANN and STL-ANN are shown in Fig. 5. Since the performance of ANN is affected by the random initialization of ANN weights (especially when dealing with regression problem), we evaluate the performance by taking average value, maximum value and minimum value from eight random initialization. As shown in Fig. 5. (a), the MFI accuracy of three modulation formats under consideration is 100% when bin number is larger than 80, no matter using STL-ANN or MTL-ANN. Fig. 5. (b) shows the OSNR estimated error versus bin number. With a small bin number, AHs cannot be described precisely. However, when bin number is too large, more data is needed to train the network. In this way, there is an optimal bin number. From Fig. 5 (b), we find the optimal bin number for MTL-ANN is 100, a half of the optimal bin number for STL-ANN. Besides, MTL-ANN suffer less performance fluctuation induced by random initialization than STL-ANN at optimal bin number.

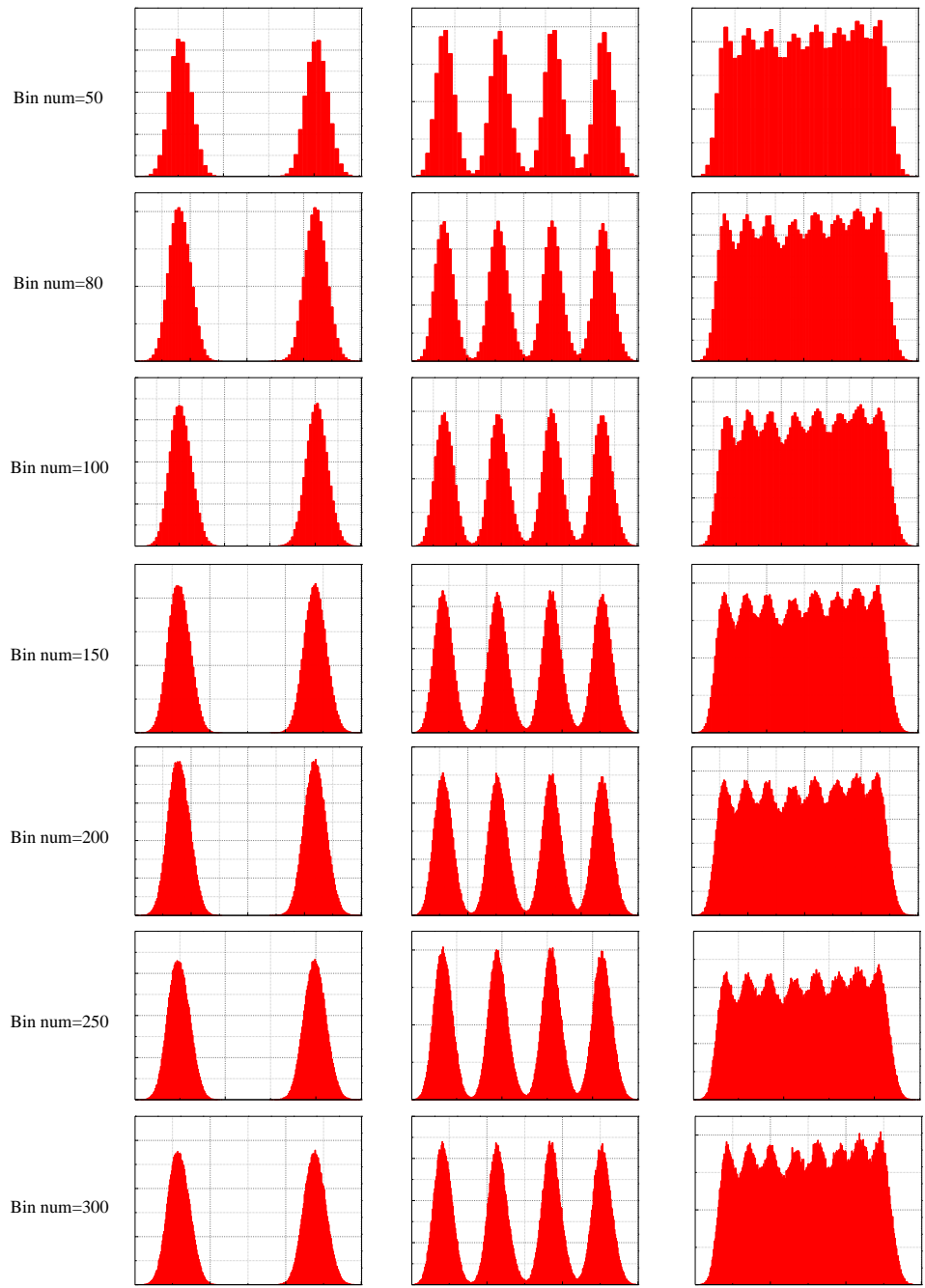

Fig. 4. AHs with different bins for three different modulation formats at 45dB OSNR. 50 bins (first row), 80 bins (second row), 100 bins (third row), 150 bins (fourth row), 200 bins (fifth row), 250 bins (sixth row) and 300 bins (seventh row).

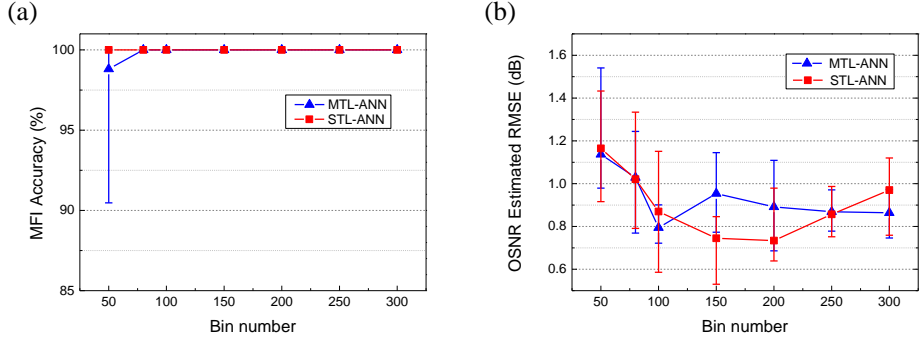

Fig. 5. (a) MFI accuracy and (b) OSNR estimated RMSE versus bin number for MTL-ANN and STL-ANN (Average, maximum and minimum value from eight random initialization).

After optimizing bin number of AHs, we investigate the optimal neuron number in shared hidden layer. The bin numbers of AHs for STL-ANN and MTL-ANN are set as 200 and 100, respectively. The neurons in specific hidden layer 1, 2 and 3 are also designed a half of the neurons in their front layer. Fig. 6 shows OSNR estimated error versus neurons in shared hidden layer for STL-ANN and MTL-ANN. The optimal neuron numbers in shared hidden layer for STL-ANN and MTL-ANN are 110 and 60, respectively. The optimal neuron number is near half of the bin number.

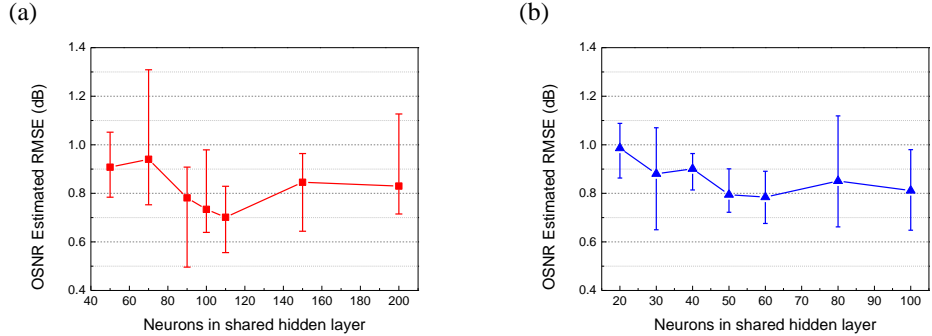

Fig. 6. OSNR estimated RMSE versus neurons in shared hidden layer for (a) STL-ANN, (b) MTL-ANN (Average, maximum and minimum value from eight random initialization).

Fig. 7 shows OSNR estimated error versus ratio of OSNR loss weight to MFI loss weight at optimal network hyperparameters. Loss weights of different tasks are defined in Fig. 2. As can be seen from Fig. 7, the OSNR estimated error decrease monotonically when the ratio is less than 7. Which means the OSNR monitoring task plays a more important role than MFI task in our designed MTL-ANN. The red dotted line represents the optimal average OSNR estimated RMSE of STL-ANN. When the loss weight ratio of OSNR to MFI is between 3 and 7, the OSNR monitoring performance of MTL-ANN is better than STL-ANN. The optimal loss weight ratio of OSNR to MFI is 5 and the RMSE of OSNR estimated error is 0.63 dB at this situation. An important point is that the neurons needed for MTL-ANN is almost the half of STL-ANN. Fig. 8 shows the estimated OSNRs versus true OSNRs of MTL-ANN at optimal bin number, shared hidden layer neurons and loss weight ratio. As can be seen form the figure, our designed MTL-ANN works well when OSNR at a low level. When OSNR is larger than 41 dB, the estimated error increase. The reason is that the quantization noise of digital-to-analog converter (DAC) makes the estimated SNR in electronic field is lower than the real SNR, especially at high OSNR situation [26].

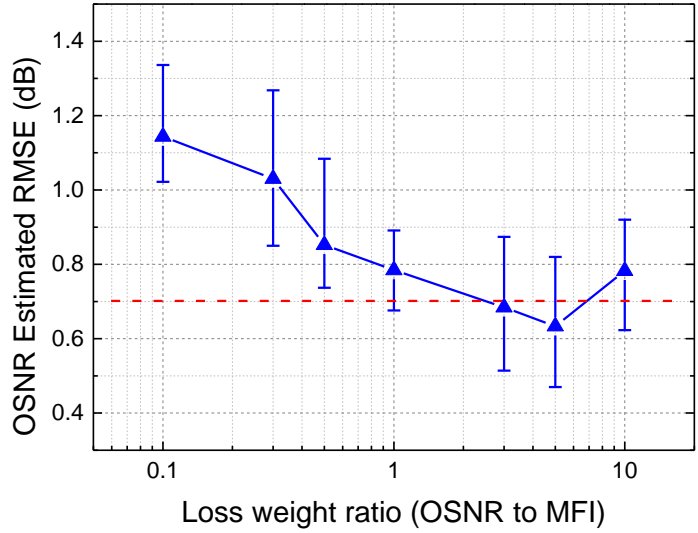

Fig. 7. OSNR estimated RMSE versus ratio of OSNR loss weight to MFI loss weight (Average, maximum and minimum value from eight random initialization).

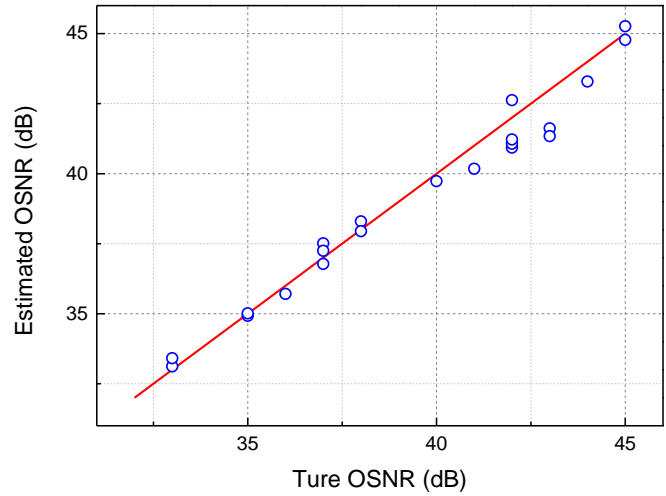

Fig. 8. True OSNRs versus estimated OSNRs of MTL-ANN.

## 5. Conclusion

In this paper, we have proposed and experimentally demonstrated an intelligent optical performance monitor using MTL-ANN to simultaneously monitoring OSNR and identify modulation format with signals' AHs. The MFI accuracy of three widely-used modulation formats in IM/DD system under consideration is 100% in estimated OSNR range (32 dB to 45 dB). Besides, OSNR monitoring with RMSE of 0.63 dB is achieved. By choosing the weighted sum of different tasks' loss function as total loss function, the effect of different tasks to the network is considered. In our work, OSNR monitoring task plays a more important role than

MFI task for designed MTL-ANN. In this intelligent optical performance monitor, only a single MTL-ANN with reduced network structure is deployed which enables reduced-complexity OPM devices. Therefore, it is attractive for real-time multi-parameters estimation in future heterogeneous optical network.

## Funding

This work was in part supported by the NSFC Program (No. 61431003, 61601049, and 61625104); BUPT Excellent Ph.D. Students Foundation (CX2018115); Fund of State Key Laboratory of Information Photonics and Optical Communications, BUPT (No. IPOC2017ZT08).